\documentclass[aps, prd, amsmath, floats, floatfix, twocolumn, nofootinbib, superscriptaddress]{revtex4-1}
\usepackage[pdftex]{graphicx}
\usepackage{subcaption}
\usepackage{color}
\usepackage{latexsym}
\usepackage[colorlinks]{hyperref}
\usepackage{amsthm,amsmath,amssymb}
\usepackage{mathrsfs}
\usepackage{booktabs}
\usepackage{multirow}

\newcommand{\be}{\begin{eqnarray}}
\newcommand{\ee}{\end{eqnarray}}

\begin{document}

\title{Testing non-circular black hole spacetime with X-ray reflection}

\author{Leda~Gao}
\affiliation{Center for Astronomy and Astrophysics, Department of Physics, Fudan University, Shanghai 200438, China}

\author{Swarnim~Shashank}
\affiliation{Institut f\"ur Astronomie und Astrophysik, Eberhard-Karls Universit\"at T\"ubingen, D-72076 T\"ubingen, Germany}
\affiliation{Center for Astronomy and Astrophysics, Department of Physics, Fudan University, Shanghai 200438, China}

\author{Cosimo~Bambi}
\email[Corresponding author: ]{bambi@fudan.edu.cn}
\affiliation{Center for Astronomy and Astrophysics, Department of Physics, Fudan University, Shanghai 200438, China}
\affiliation{School of Natural Sciences and Humanities, New Uzbekistan University, Tashkent 100007, Uzbekistan}

\begin{abstract}
X-ray reflection spectroscopy is a powerful tool for testing the Kerr hypothesis and probing the strong gravity regime around accreting black holes. Most tests of General Relativity (GR) assume that the spacetime around a black hole is circular, meaning the metric possesses a specific symmetry structure common to the Kerr solution. However, deviations from circularity are predicted by various modified gravity theories and non-vacuum General Relativity solutions. In this work, we test a specific non-circular metric constructed based on a locality principle, where the deviation from the Kerr spacetime is driven by the local spacetime curvature. To accurately model the reflection spectrum in this background, we implement a relativistic ray-tracing code in horizon-penetrating (ingoing Kerr) coordinates, which are favored for their ability to avoid introducing curvature singularities at the horizon in non-circular spacetimes. We apply this model to the high-quality \textit{NuSTAR} spectrum of the Galactic black hole binary EXO 1846--031. Our spectral analysis reveals a source with a high inclination angle ($\iota \approx 76^{\circ}$) and a near-extremal spin parameter ($a_* \approx 0.98$). While we identify a global minimum in the parameter space suggesting a non-zero deformation ($\ell_{\mathrm{NP}} \approx 0.12$), the 99\% confidence interval fully encompasses the Kerr limit ($\ell_{\mathrm{NP}}=0$). We conclude that the current X-ray reflection data for EXO 1846--031 are consistent with the Kerr hypothesis. This work demonstrates the feasibility of using X-ray reflection spectroscopy to constrain non-circular metrics and establishes a framework for future tests.
\end{abstract}

\maketitle


\section{Introduction}

As one of the most successful theories in physics, General Relativity (GR), proposed by Einstein, has predicted and explained a wide range of impressive phenomena, such as the expansion of the Universe, the existence of black holes, and gravitational waves. Thanks to rapid technological developments over the last several decades, we have detected and tested many of these phenomena, notably gravitational waves through the LIGO-VIRGO-KAGRA (LVK) Collaboration~\cite{LIGOScientific:2016lio,LIGOScientific:2021sio,LIGOScientific:2025rid} and black hole images taken by the Event Horizon Telescope Collaboration~\cite{EventHorizonTelescope:2019dse}. All tests performed so far support the correctness of GR and place constraints on possible theories beyond GR. From a theoretical perspective, however, research into theories beyond GR continues, as it is widely believed that GR may not be the ultimate theory of gravity. Reasons include the presence of spacetime singularities in most physically relevant solutions of the Einstein Equations, the puzzle of dark energy in cosmological models, etc. One avenue to search for deviations from GR is to test for potential deviations from the Kerr metric~\cite{2017RvMP...89b5001B,Bambi_book,2016CQGra..33e4001Y} a stationary and axisymmetric vacuum solution to the Einstein field equations, the Kerr metric describes a rotating black hole, an object whose existence in the Universe is now well-established. Therefore, Kerr black holes are excellent candidates for probing physics beyond our current knowledge.

There are generally two methods to introduce deviations from the Kerr metric. One is to derive a rotating black hole solution from a specific modified gravity theory. However, due to the high complexity of the field equations, there are few non-trivial exact solutions in modified gravity theories available for investigation, such as the disformal Kerr solution in generalised scalar–tensor
gravity~\cite{Anson:2020trg,BenAchour:2020fgy,Minamitsuji:2020jvf}. As an alternative approach, one can describe deviations from the Kerr metric agnostically. Based on physical intuition, one can parameterize the deviation from the Kerr metric to construct a phenomenological metric. This allows us to search for ``smoking guns'' of GR modifications. 
The majority of deformed Kerr metrics proposed in the literature are circular, like the original Kerr metric. As defined by the condition of circularity, there exists a foliation of two-dimensional surfaces everywhere orthogonal to the Killing vectors. Therefore, for a circular spacetime, one can choose coordinates such that the metric is block diagonal; the final metric has only five non-vanishing components. For the Kerr metric in Boyer-Lindquist coordinates, the only non-vanishing off-diagonal component is $g_{t\phi}$.  
However, circularity conditions are not necessarily satisfied in some modified gravity theories, such as disformed Kerr solution~\cite{Anson:2020trg}, Einstein-\ae ther theory~\cite{Adam:2021vsk}, and Generalized Proca theories~\cite{Fernandes:2026rjs}, or even in non-vacuum GR solutions~\cite{Gourgoulhon1993N}.  
As stated in Ref.~\cite{Babichev:2025szb}, it is meaningful to construct stationary and axisymmetric metrics without assuming circularity and to investigate the breaking of this condition.

Refs.~\cite{Delaporte:2022acp,Babichev:2025szb} discuss the general form of metrics that break circularity. A relatively general parameterization of non-circular metrics may contain several parameters which can depend on the coordinates $r$ and $\theta$~\cite{Delaporte:2022acp,Ghosh:2024arw}. This introduces significant challenges when testing the model with observational data. First, the signal quality may not be sufficient to constrain multiple parameters. Second, a complex metric can make the construction of observable templates computationally expensive. Therefore, although it may introduce some loss of generality, a non-circular metric with a simple form is a good candidate for identifying signatures of circularity violation. Refs.~\cite{Eichhorn:2021iwq,Eichhorn:2021etc} proposed a family of metrics based on a locality principle, which states that the deviation induced by new physics is tied to local curvature scales. This metric substitutes the black hole mass parameter $M$ with a mass function $M(r,\theta)$, which is a function of a curvature invariant. This type of deformation has been proven to be non-circular~\cite{Delaporte:2022acp,Babichev:2025szb}. In this work, we test this type of non-circular metric using X-ray reflection spectroscopy, a powerful technique for probing the strong gravity regions of accreting black holes~\cite{Bambi:2023hez}, both stellar-mass and supermassive.

In the disk-corona model (Fig.~\ref{fig: corona}), a black hole accretes from a geometrically thin and optically thick accretion disk. Thermal photons emitted from the disk can inverse Compton scatter off free electrons in the corona. The Comptonized photons from the corona can illuminate the accretion disk and produce reflection photons. The reflection spectrum in the rest frame of the disk material is characterized by narrow fluorescent emission lines below $10\,\mathrm{keV}$, the Fe K-edge at $7$--$10\,\mathrm{keV}$, and a Compton hump with a peak around $20$--$40\,\mathrm{keV}$~\cite{Fabian:1995qz,Zoghbi_2010,Risaliti:2013cga,Ross:2005dm,2010ApJ...718..695G}. Since the emission of the reflection spectrum can originate close to the innermost stable circular orbit (ISCO), relativistic effects must be fully considered to model the final spectrum detected by a distant observer. Refs.~\cite{Bambi:2016sac,Abdikamalov:2019yrr,2020ApJ...899...80A} extended the relativistic ray-tracing in the standard reflection package \texttt{relxill} to non-Kerr spacetimes. This package, \texttt{relxill\_nk}, has been successfully applied to observational data to constrain theories beyond GR, such as conformal gravity~\cite{Zhou:2018bxk,Zhou:2019hqk}, Kaluza-Klein gravity~\cite{Zhu:2020cfn}, asymptotically safe quantum gravity~\cite{Zhou:2020eth}, Einstein-Maxwell dilaton-axion gravity~\cite{Tripathi:2021rwb}, and bumblebee gravity~\cite{Gu:2022grg}(see also, e.g., Refs.~\cite{Tao:2023hou,Liao:2024ezc}). Depending on the quality of the observed data and the specific black hole metric, X-ray reflection spectroscopy may provide constraints comparable to, or even more stringent than, other major techniques for testing GR, such as gravitational waves and black hole imaging~\cite{2019ApJ...874..135T,Tripathi:2020yts,Tripathi:2021rqs,Riaz:2022rlx,2024arXiv240108545T,Zhao:2025mwq}. 
These relativistic ray-tracing codes for testing non-Kerr metrics were originally designed for Boyer-Lindquist coordinates. However, Refs.~\cite{Delaporte:2022acp,Babichev:2025szb} suggest that metrics in horizon-penetrating coordinates, such as ingoing Kerr coordinates, may be preferred for parameterizing non-circular metrics deviating from Kerr spacetime. One reason is that it is easier to avoid introducing curvature singularities at the horizon in horizon-penetrating coordinates. Therefore, we modified the traditional ray-tracing code to work in ingoing Kerr coordinates, which will facilitate future testing of non-circular metrics.

The outline of this paper is as follows. Section~\ref{subsec: non-circular_metric} presents the definition of non-circularity and the specific non-circular metric used in this work. In Section~\ref{sec: X-ray_reflection_spectroscopy}, we discuss the details of X-ray reflection spectroscopy, including the different components of the electromagnetic spectrum, the construction of the spectral model for fitting, and the use of ray-tracing to model relativistic effects. Furthermore, we discuss the broadened single iron lines computed in the non-circular metric. In Section~\ref{sec:observationa_constraints}, observational constraints are placed on the deformation parameter through the analysis of the Galactic black hole binary EXO 1846--031. Section~\ref{sec:conclusion} presents the conclusions. In Appendix~\ref{app: ray-tracing_ingoing}, we derive and present the key equations necessary for implementing ray-tracing in ingoing Kerr coordinates. We employ natural units in which $c = G_{\rm N} = 1$ throughout the manuscript.

\section{Choice of Non-Circular Metric}\label{subsec: non-circular_metric}

By utilizing the definition in Ref.~\cite{Babichev:2025szb}, a spacetime is said to be circular if the two-dimensional surfaces generated by the action of the Killing isometries are everywhere orthogonal to a family of codimension-two surfaces. Specifically, this requirement can be translated into the following mathematical expression using Frobenius’ theorem. If a spacetime is circular, the following conditions must hold everywhere:
\begin{subequations} \label{eqn:original_circular_condition}
\begin{align}
(\xi_\mu \mathrm{d}x^\mu) \wedge (\psi_\nu \mathrm{d}x^\nu) \wedge \mathrm{d} (\xi_\rho \mathrm{d}x^\rho) &= 0, \\
(\psi_\mu \mathrm{d}x^\mu) \wedge (\xi_\nu \mathrm{d}x^\nu) \wedge \mathrm{d} (\psi_\rho \mathrm{d}x^\rho) &= 0, 
\end{align}
\end{subequations}
where $\xi^\mu$ and $\psi^\mu$ are the Killing vectors associated with stationarity and axisymmetry, respectively.

In Ref.~\cite{Babichev:2025szb}, by making gauge choices for the metric of stationary and axially symmetric spacetimes, a general form of the metric is written as 
\begin{equation}\label{eqn:Kerr_like_gauge}
\begin{split}
    \mathrm{d}s^2 &= g_{vv} \mathrm{d}v^2 + g_{\theta\theta} \mathrm{d}\theta^2 + g_{rr} \mathrm{d}r^2 + g_{\phi\phi} \mathrm{d}\phi^2 \\
         &\quad + 2g_{v\phi} \mathrm{d}v \mathrm{d}\phi + 2g_{vr} \mathrm{d}v \mathrm{d}r + 2g_{r\phi} \mathrm{d}r \mathrm{d}\phi, 
\end{split}
\end{equation}
where the metric components depend only on $r$ and $\theta$. 
We list the key properties here; detailed descriptions for obtaining this metric can be found in the original paper:
\begin{itemize}
    \item There are two coordinates $v$ and $\phi$ associated with two Killing vectors $\xi^\mu$ and $\psi^{\mu}$, such that $\xi^\mu\partial_\mu = \partial_v$ and $\psi^\mu\partial_{\mu}=\partial_\phi$.
    \item \textit{Orthogonal gauge}: The coordinate basis vector $\partial_{\theta}$ is orthogonal to $\partial_{v}$, $\partial_{r}$, and $\partial_{\phi}$, which implies that the metric components satisfy $g_{\mu\theta}\propto \delta^{\theta}_{\ \mu}$. 
    \item \textit{Kerr-like gauge}: The metric reduces to a Kerr-like form by setting $g_{rr} = 0$.  
\end{itemize}
By solving the circularity conditions \eqref{eqn:original_circular_condition} for the metric \eqref{eqn:Kerr_like_gauge}, the necessary and sufficient conditions for this metric to be circular are\footnote{The derivation does not require the metric \eqref{eqn:Kerr_like_gauge} to be in the Kerr-like gauge with $g_{rr} = 0$.}
\begin{subequations}\label{eqn:derived_circular_condition}
\begin{align}
 g^{vr} &= g^{rr}f(r), \\
 g^{r\phi} &= g^{rr}h(r).
\end{align}
\end{subequations}
One method to construct a non-circular metric that violates the circularity conditions in \eqref{eqn:derived_circular_condition} is to replace the mass parameter $M$ in the Kerr metric with a mass function $m(r,\theta)$ that depends non-trivially on the angle $\theta$~\cite{Babichev:2025szb}. In this case, the deformed Kerr metric is expressed in ingoing Kerr coordinates as
\begin{equation}\label{eqn:non_circular_metric}
\begin{split}
    \mathrm{d}s^2 &= -\left( 1 - \frac{2rM(r,\theta)}{\Sigma} \right) \mathrm{d}v^2 + \Sigma \, \mathrm{d}\theta^2 + \frac{\tilde{A} \sin^2 \theta}{\Sigma} \mathrm{d}\phi^2 \\
    &\quad + 2 \, \mathrm{d}v \, \mathrm{d}r - 2a \sin^2 \theta \, \mathrm{d}r \, \mathrm{d}\phi \\
    &\quad - \frac{4M(r,\theta)ar \sin^2 \theta}{\Sigma} \mathrm{d}v \, \mathrm{d}\phi,
\end{split}
\end{equation}
where $\tilde{\Delta} := r^2 - 2rM(r,\theta) + a^2$, $\Sigma := r^2 + a^2 \cos^2 \theta$, and $\tilde{A} := (r^2 + a^2)^2 - \tilde{\Delta} a^2 \sin^2 \theta$.

However, Ref.~\cite{Babichev:2025szb} does not provide a specific form for $M(r,\theta)$. By searching through the literature, we selected a generalization of the Kerr metric proposed in Refs.~\cite{Delaporte:2022acp,Eichhorn:2021iwq,Eichhorn:2021etc} based on a locality principle. Briefly, the deviation of the generalized metric from the Kerr spacetime depends on the local curvature. The metric is designed to deviate more strongly from the Kerr metric in regions with larger local curvature. The local curvature for a Kerr black hole with mass $m$ and spin parameter $a$ is characterized by the invariant $K_\textrm{GR}$,
\begin{equation}\label{eqn:KGR}
    K_\textrm{GR} = \frac{48m^2}{(r^2+a^2\cos^2\theta)^3}. 
\end{equation}
The mass function takes the form $M(K_\textrm{GR})=M(r,\theta)$ and can be chosen freely as long as it satisfies the correct asymptotic limits, $M(K_\textrm{GR}\rightarrow0)\rightarrow m$ and $M(K_\textrm{GR}\rightarrow\infty)\rightarrow 0$, and falls off faster than $\mathcal{O}((K_\textrm{GR}\ell_{\textrm{NP}}^4)^{-1/2})$ between the limits, where $\ell_{\textrm{NP}}$ is the deformation parameter characterizing the deviation from the regular Kerr spacetime. As one might expect, there exists a class of mass functions satisfying these requirements. Several choices of mass functions are designed in~\cite{Eichhorn:2021iwq,Eichhorn:2021etc}. In this work, we choose the mass parameter $M(r,\theta)$ discussed thoroughly in~\cite{Eichhorn:2021iwq,Eichhorn:2021etc}, given as\footnote{We note that this rotating black hole metric can be obtained from the non-rotating metric and applying the Newman-Janis algorithm~\cite{Newman:1965tw}. We need to consider the complex transformation used in the Schwarzschild/Kerr case; the invariant $K_\textrm{GR}$ transforms as $K_\textrm{GR} = 48 m^2/r^6 \rightarrow 48 m^2/\Sigma^3$~\cite{Bambi:2013ufa}.}  
\begin{equation}\label{eqn:non_circular_mass}
    M(r,\theta) = \frac{m}{1+(\ell_{\textrm{NP}}^4K_\textrm{GR})^{\frac{\beta}{2}}},
\end{equation}
where $\beta > 1$ ensures a completely well-defined geometry.
Here we choose $\beta = 2$, the same value used in~\cite{Eichhorn:2021etc}, as it is the integer value leading to the largest deformation effect. 
Although this work cannot be viewed as a general test for non-circular metrics, the analysis framework developed here can be easily applied to any other non-circular Kerr-like metric in ingoing Kerr coordinates. 

\begin{figure*}[t]
\centering
\includegraphics[width=0.44\linewidth]{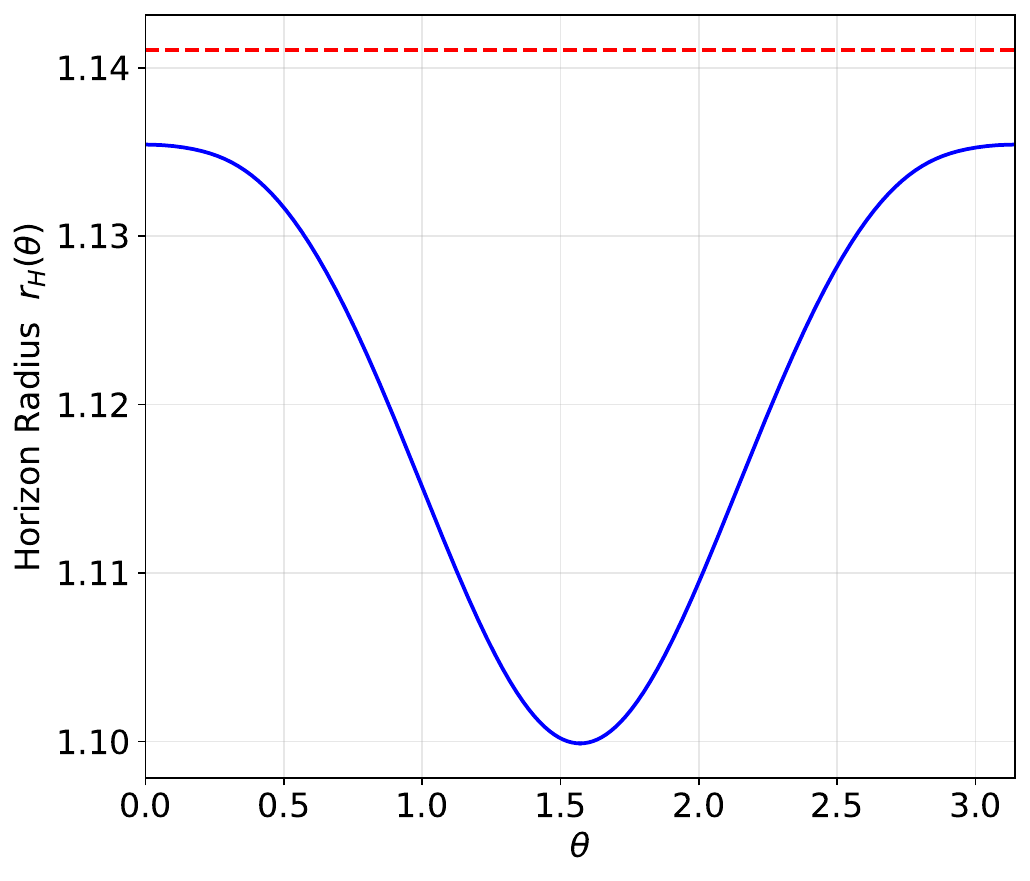}
\vspace{-0.2cm}
\includegraphics[width=0.53\linewidth]{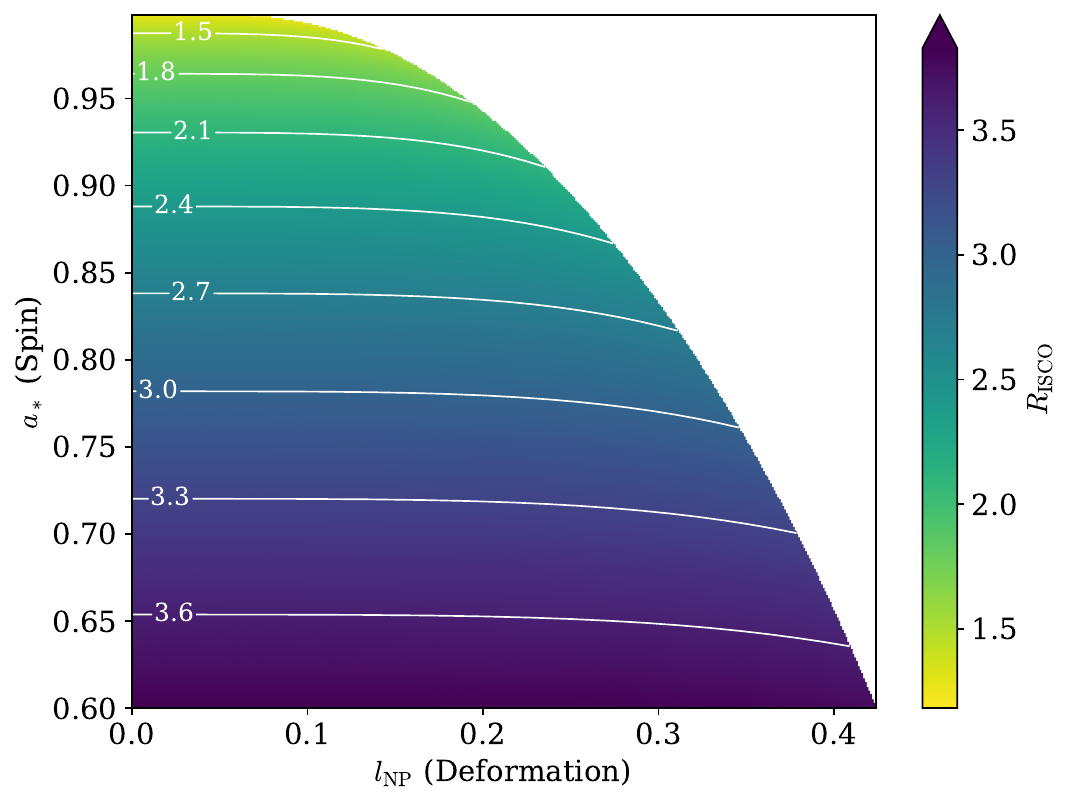}
\vspace{-0.2cm}
\caption{Left panel: The horizon radius $r_H(\theta)$ is plotted as a function of the polar angle $\theta$ as a solid blue line for a spin parameter $a_*=0.99$ and deformation parameter $\ell_{\textrm{NP}}=0.1137$ (the maximum $\ell_{\textrm{NP}}$), while the horizon radius for a Kerr black hole is plotted as a red dashed line for comparison. Right panel: Contour levels for the values of the ISCO radius $R_\mathrm{ISCO}$ as a function of spin parameter $a_*$ and deformation parameter $\ell_{\textrm{NP}}$. The white region is excluded from our study because the spacetime possesses a naked singularity for these parameter combinations.}
\label{fig:isco_non_circular}
\end{figure*}

To find the location of the horizon, $r_H = H(\theta)$, the following ordinary differential equation 
\begin{equation}\label{eqn:Horizon_condition}
g^{rr}(r=H(\theta))+g^{\theta\theta}(r=H(\theta))\left(\frac{\mathrm{d}H}{\mathrm{d}\theta}\right)^2=0
\end{equation}
must be solved numerically~\cite{Eichhorn:2021iwq}. To obtain the initial condition for solving Eq.~\eqref{eqn:Horizon_condition}, the horizon condition at $\theta=\pi/2$ is chosen to be $g^{rr}=0$, which ensures the horizon possesses axisymmetry and reflection symmetry about the equatorial plane. The initial radius $r_H(\pi/2)$ satisfies 
\begin{equation} \label{eqn:horizon_initial}
    r_H^2+a^2-2M(r,\theta)r_H =0. 
\end{equation}
Furthermore, we require that $H'(\theta)=0$ at $\theta=0$ and $\pi$ to ensure a smooth horizon surface. 
This condition and the existence of a root for Eq.~\eqref{eqn:horizon_initial} set constraints on the deformation parameter $\ell_{\textrm{NP}}$ to ensure the existence of a smooth outer horizon surface for the spacetime\footnote{The condition $H'(\theta)=0$ at $\theta=0$ and $\pi$ can only be tested numerically by integrating $r_H$ from $\theta=\pi/2$ to $0$ or $\pi$. During calculation, we found that this condition can be violated for the extremal $\ell_{\textrm{NP}}$ obtained by only considering the existence of a root for Eq.~\eqref{eqn:horizon_initial}. Therefore, we consider a maximum $\ell_{\textrm{NP}}$ slightly smaller than that extremal $\ell_{\textrm{NP}}$ when constructing the parameter space.}. 
Then we integrate Eq.~\eqref{eqn:Horizon_condition} to $\theta=0$ or $\theta=\pi$ using the initial condition $r_H = r_H(\pi/2)$ and $\frac{\mathrm{d}H}{\mathrm{d}\theta}|_{\theta=\pi/2}= 0$. 

As an example, in the left panel of Fig.~\ref{fig:isco_non_circular}, the horizon radius $r_H(\theta)$ is plotted as a function of the polar angle $\theta$ as a solid blue line for a black hole with dimensionless spin parameter\footnote{During the computation, we often consider the dimensionless spin parameter $a_*=a/M$.} $a_*=0.99$ and deformation parameter $\ell_{\textrm{NP}}=0.1137$, which is the maximum $\ell_{\textrm{NP}}$, while the horizon radius for a Kerr black hole is plotted as a red dashed line for comparison. The location of the horizon shows the most deformation on the equatorial plane and the least deformation at the poles. 

The ISCO in the equatorial plane is crucial for constructing the model of thin accretion disks and analyzing the reflection spectrum of the black hole (the method for calculating the ISCO radius can be found in~\cite{Bambi:2024hhi}). The right panel of Fig.~\ref{fig:isco_non_circular} shows the value of the ISCO radius in the spacetime of Eq.~\eqref{eqn:non_circular_metric} as a function of the black hole spin parameter $a_*$ and the deformation parameter $\ell_{\textrm{NP}}$. The white region is the part of the parameter space where a smooth horizon surface does not exist. For black holes with higher spin, the range of $\ell_{\textrm{NP}}$ ensuring the existence of a horizon is smaller. In the limit of $a_*\rightarrow1$, $\ell_{\textrm{NP}}\rightarrow0$.  
Similar to the decrease in the horizon radius for the deformed spacetime, the ISCO radius tends to decrease as $\ell_{\textrm{NP}}$ increases for a specific spin value.

\section{X-RAY REFLECTION SPECTROSCOPY}\label{sec: X-ray_reflection_spectroscopy}

\begin{figure}[t]
\centering
\includegraphics[width=0.9\linewidth]{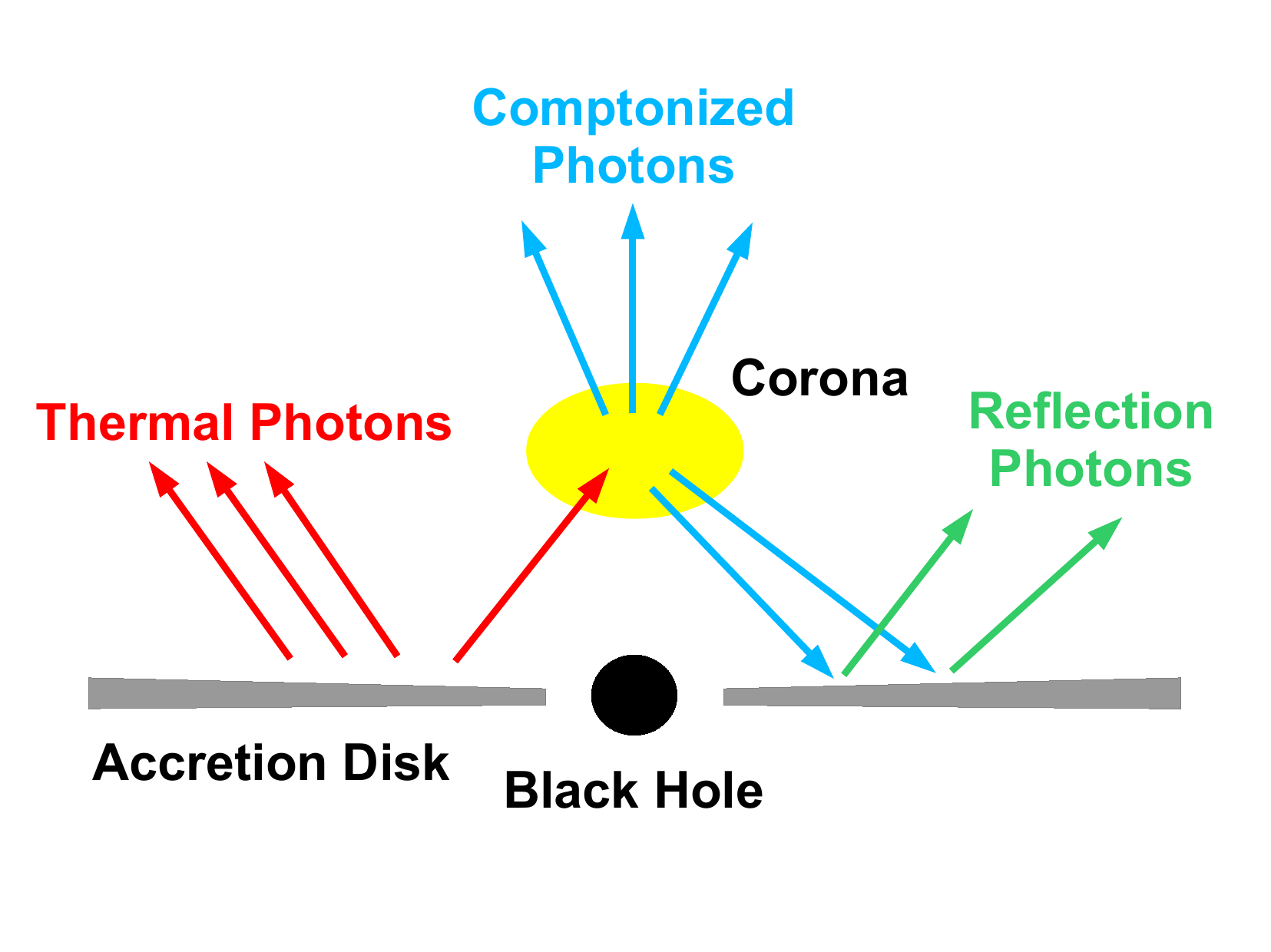}
\vspace{-0.2cm}
\caption{This diagram shows each component of the electromagnetic spectrum in the disk-corona model: thermal emission from the cold disk (red arrows), a Comptonized spectrum from the hot corona (blue arrows), and a reflection spectrum from the disk (green arrows). Figure from Ref.~\cite{Bambi:2024hhi}.
\label{fig: corona}}
\end{figure}

X-ray reflection spectroscopy is a well-established and powerful tool for testing models of accreting black holes and the physics in strong gravitational fields~\cite{Bambi:2020jpe}. In this work, the accreting black hole is described by the disk-corona model. The accretion disk is cold, geometrically thin, and optically thick, and is described by the Novikov-Thorne model~\cite{1973blho.conf..343N,Page:1974he}. Each point on the surface of the accretion disk is assumed to be in local thermal equilibrium and emits a blackbody-like spectrum. Therefore, the entire accretion disk emits a multi-temperature blackbody spectrum. The corona is a hot electron plasma ($T_e\sim100\,\mathrm{keV}$) located near the black hole and the inner part of the accretion disk. However, the exact morphology of the corona is not yet fully understood. There are several different models, such as the lamppost model, the sandwich model, and spherical or toroidal models. The geometry of the corona can change, and different coronal geometries may coexist. A schematic diagram illustrating the disk-corona model and the X-ray spectrum of an accreting black hole is shown in Fig.~\ref{fig: corona}.\footnote{Note that we use a lamppost model to illustrate the morphology of the corona in Fig.~\ref{fig: corona}. However, we do not specify the morphology of the corona in this work. Instead, we use a phenomenological model for the Comptonized spectrum from the corona.}

Since the disk is cold and the corona is hot, thermal photons emitted from the disk can inverse Compton scatter off the free electrons in the corona. A fraction of the Comptonized photons from the corona (blue arrows in Fig.~\ref{fig: corona}) illuminates the disk. The reflection spectrum (green arrows in Fig.~\ref{fig: corona}) is produced by fluorescent emission, as Compton scattering and absorption in the disk result from the irradiation by Comptonized photons. 
The reflection spectrum in the rest frame of the disk material is characterized by narrow fluorescent emission lines below $10\,\mathrm{keV}$, the Fe K-edge at $7$--$10\,\mathrm{keV}$, and a Compton hump with a peak around $20$--$40\,\mathrm{keV}$. To accurately model the reflection spectrum of the entire accretion disk as seen by a distant observer, we must take into account all relativistic effects, such as Doppler boosting and gravitational redshift. By incorporating the non-circular spacetime into the relativistic model, the test of the non-circular metric can be achieved through the analysis of the relativistically blurred reflection spectrum (X-ray reflection spectroscopy). We summarize the model used to describe the X-ray spectrum in Sec.~\ref{subsec: X-ray_spectrum} and the construction of the relativistic model in Sec.~\ref{subsec: relativistic_model}.

\subsection{Modeling of X-ray spectrum}\label{subsec: X-ray_spectrum}

As illustrated in Fig.~\ref{fig: corona}, the X-ray spectrum emitted from an accreting black hole can generally be decomposed into three components: the thermal emission from the disk, the direct emission of Comptonized photons from the corona, and the reflection spectrum from the disk. In this work, we construct a spectral model following the configuration of Model 1 in Ref.~\cite{Li:2024eue}.

The spectrum of the Comptonized photons from the corona is typically well-approximated by a power law with low- and high-energy cutoffs. We employ the thermal Comptonization model \texttt{nthComp}~\cite{Zdziarski:1996wq,Zycki:1999cm} to describe this continuum component. This model is parameterized by the photon index $\Gamma$ and the electron temperature $kT_e$. The \texttt{diskbb} model~\cite{Pringle:1981adi,Mitsuda:1984nv} is used to describe the direct thermal emission from the accretion disk parameterized by the temperature at the inner edge of the disk ($T_\mathrm{in}$) and the normalization of the component.

The reflection spectrum is modeled using \texttt{relxillionCp\_nk}~\cite{Abdikamalov:2021rty}. This model is a variant of the \texttt{relxill\_nk} package~\cite{Bambi:2016sac,Abdikamalov:2019yrr}. The \texttt{relxill\_nk} package extends the relativistic ray-tracing of the standard \texttt{relxill} package~\cite{Dauser:2013xv,Garcia:2013oma,Garcia:2013lxa} to non-Kerr spacetimes. Compared to \texttt{relxill\_nk}, \texttt{relxillionCp\_nk} introduces two specific features.
The first feature of \texttt{relxillionCp\_nk} is the implementation of an empirical power-law form for the radial profile of the ionization parameter:
\begin{equation}\label{eqn:ionization_parameter}
    \xi(r) = \xi_0\left(\frac{R_\mathrm{in}}{r}\right)^{\alpha_\xi},
\end{equation}
where $\xi_0$ is the ionization parameter at the inner edge of the disk, $R_\mathrm{in}$ is the radial coordinate of the inner edge, and $\alpha_\xi$ is the power-law index of the profile. Previous studies of our target source, the black hole binary EXO 1846-031, have shown that a model with a radial ionization gradient provides an improved fit~\cite{Abdikamalov:2021rty}. Although this feature has only a modest impact on the estimation of the black hole spin and the constraints on the deformation parameter, we include it to ensure the highest possible fitting quality.

The second feature of \texttt{relxillionCp\_nk} is the use of the \texttt{nthcomp} Comptonization continuum as the incident spectrum (the illuminating radiation). In our analysis, we tie the photon index $\Gamma$ and electron temperature $kT_e$ of the incident spectrum in \texttt{relxillionCp\_nk} to the corresponding parameters of the direct continuum model \texttt{nthcomp}.

Finally, we use \texttt{tbabs} to account for Galactic absorption, adopting the interstellar medium abundance table from Ref.~\cite{Wilms:2000ez}. In XSPEC notation, our total model is defined as: $\texttt{constant*tbabs*(diskbb+nthcomp+relxillionCp\_nk)}$.
To constrain the deformation parameter $\ell_{\mathrm{NP}}$ for the non-circular metric described in Sec.~\ref{subsec: non-circular_metric}, \texttt{relxillionCp\_nk} requires Flexible Image Transport System (FITS) files containing the transfer functions specific to that metric. The calculation of these transfer functions is detailed in the following section.

\subsection{Modeling of relativistic effect} \label{subsec: relativistic_model}

To model the relativistic effects and analyze the reflection spectrum, the transfer function for geometrically thin and optically thick accretion disks is computed numerically for the non-circular metric~\cite{Cunningham:1975zz,SPEITH1995109}. 
The observed flux of a thin accretion disk seen by a distant observer can be written as 
\begin{equation}\label{eqn:observed_flux}
\begin{split}
F_{\mathrm{o}}(\nu_{\mathrm{o}}) &= \int I_{\mathrm{o}}(\nu_{\mathrm{o}}, X, Y) \, \mathrm{d}\Omega \\
&= \frac{1}{D^2}\int g^3 I_{\mathrm{e}}(\nu_{\mathrm{e}}, r_{\mathrm{e}}, \vartheta_{\mathrm{e}}) \, \mathrm{d}X \, \mathrm{d}Y, 
\end{split}
\end{equation}
where $I_{\mathrm{o}}$ and $I_{\mathrm{e}}$ are the specific intensities of the radiation measured by the distant observer and in the rest frame of the emitting material, respectively. $\nu_{\mathrm{o}}$ and $\nu_{\mathrm{e}}$ are the photon frequencies measured by the distant observer and the emitter, respectively. $r_{\mathrm{e}}$ is the emission radius in the accretion disk, and $\vartheta_{\mathrm{e}}$ is the emission angle in the rest frame of the disk material. 
$\mathrm{d}\Omega=\mathrm{d}X\mathrm{d}Y/D^2$ is the infinitesimal solid angle subtended by the source on the sky of the distant observer, where $X$ and $Y$ are the Cartesian coordinates in the plane of the distant observer, and $D$ is the distance between the observer and the source. $g=\nu_{\mathrm{o}}/\nu_{\mathrm{e}}$ is the redshift factor, and the relation $I_{\mathrm{o}}=g^3 I_{\mathrm{e}}$ follows from Liouville's theorem.

In X-ray spectral analysis, it is necessary to rapidly generate many spectral templates across the full parameter space to identify the best-fit model. A widely used method is to recast Eq.~\eqref{eqn:observed_flux} in terms of Cunningham's transfer function~\cite{Cunningham:1975zz}, yielding the following expression:
\begin{equation}\label{eqn:transfer_function}
\begin{split}
F_{\mathrm{o}}(\nu_{\mathrm{o}}) = \frac{1}{D^2} \sum_{i=1}^{2} \int_{r_{\mathrm{in}}}^{r_{\mathrm{out}}} \mathrm{d}r_{\mathrm{e}} \int_{0}^{1} \mathrm{d}g^* \, \frac{\pi r_{\mathrm{e}} g^2}{\sqrt{g^* (1-g^*)}} \\
\times f^{(i)}(g^*, r_{\mathrm{e}}, \iota) I_{\mathrm{e}}(\nu_{\mathrm{e}}, r_{\mathrm{e}}, \vartheta_{\mathrm{e}}^{(i)}),
\end{split} 
\end{equation}
where $f$ is Cunningham's transfer function, given by
\begin{equation}
f^{(i)}(g^*, r_{\mathrm{e}}, \iota) = \frac{g \sqrt{g^* (1-g^*)}}{\pi r_{\mathrm{e}}} \left| \frac{\partial(X, Y)}{\partial(r_{\mathrm{e}}, g^*)} \right|,
\end{equation}
and $|\partial(X,Y)/\partial(r_{\mathrm{e}},g^*)|$ is the Jacobian of the transformation between the Cartesian coordinates $(X,Y)$ of the distant observer and the coordinates $(r_{\mathrm{e}}, g^*)$ on the accretion disk. 
Here, $g^*$ is the relative redshift factor defined as $g^*=(g-g_{\min})/(g_{\max}-g_{\min})$, where $g_{\min} = g_{\min}(r_{\mathrm{e}},\iota)$ and $g_{\max} = g_{\max}(r_{\mathrm{e}},\iota)$ are functions of the emission radius $r_{\mathrm{e}}$ and the inclination angle of the disk $\iota$ (the angle between the disk normal and the line of sight). In Eq.~\eqref{eqn:transfer_function}, the flux integral is split into two parts involving transfer functions $f^{(1)}$ and $f^{(2)}$. This separation arises because the points on the accretion disk at the same emission radius form closed loops in the observer's image plane; the points of minimum and maximum redshift divide these loops into two distinct branches. 

To obtain the transfer function, the mapping between the coordinates $(X,Y)$ and $(r_{\mathrm{e}}, g^*)$ is calculated using a ray-tracing code. The transfer function is then tabulated in a FITS file on a grid of emission radii and relative redshifts $(r_{\mathrm{e}},g^*)$ for each point in the parameter space spanned by the black hole spin parameter $a_*$, the deformation parameter $\ell_{\mathrm{NP}}$, and the inclination angle $\iota$. Finally, this FITS file is used to construct the \texttt{relxillionCp\_nk} model. 
The entire process is described in detail in Refs.~\cite{Bambi:2016sac,Abdikamalov:2019yrr,Bambi:2024hhi}.  

In previous works~\cite{Bambi:2016sac,Abdikamalov:2019yrr}, the ray-tracing code for testing non-Kerr metrics was designed using Boyer-Lindquist coordinates. However, as discussed in Refs.~\cite{Delaporte:2022acp,Babichev:2025szb}, horizon-penetrating coordinates—such as the ingoing Kerr coordinates used here—are favored when constructing non-circular metrics. This is primarily because it is easier to avoid introducing curvature singularities at the horizon in these coordinates. As noted in Ref.~\cite{Eichhorn:2021iwq}, the metric given by Eqs.~\eqref{eqn:non_circular_metric} and~\eqref{eqn:non_circular_mass} is well-approximated by the Kerr spacetime at large distances and can be transformed to Boyer-Lindquist coordinates for a distant observer as $r\rightarrow\infty$. However, an analytical coordinate transformation to convert the entire metric into a Boyer-Lindquist form has not yet been derived. Therefore, a practical approach to implementing ray tracing is to generate photons with initial conditions in the Kerr metric using Boyer-Lindquist coordinates (representing the distant observer), and then convert these to ingoing Kerr coordinates to evolve the photons within the deformed metric. The details regarding the generation of photon initial conditions and their geodesic evolution are explained in Appendix~\ref{app: ray-tracing_ingoing}. 
All code is publicly available on GitHub\footnote{\href{https://github.com/LEDA-GAO/Non-Circular-Raytracing}{https://github.com/LEDA-GAO/Non-Circular-Raytracing}. Portions of the code are adapted from \texttt{blackray} and \texttt{raytransfer}, available at \href{https://github.com/ABHModels/}{https://github.com/ABHModels/}.}.

\subsection{Single line shapes in non-circular spacetime}\label{subsec: single_line_shapes}

In this section, assuming that the emission is a narrow iron line at $6.4\,\mathrm{keV}$, we calculate the broadened lines including relativistic effects using our ray-tracing code with the metric in Eqs.~\eqref{eqn:non_circular_metric} and~\eqref{eqn:non_circular_mass}. The local spectrum $I_{\mathrm{e}}$ is modeled as a power law with an emissivity index equal to 3, i.e., $I_{\mathrm{e}} \propto 1/r_{\mathrm{e}}^3$. 

\begin{figure*}[t]
\centering
\includegraphics[width=0.9\linewidth]{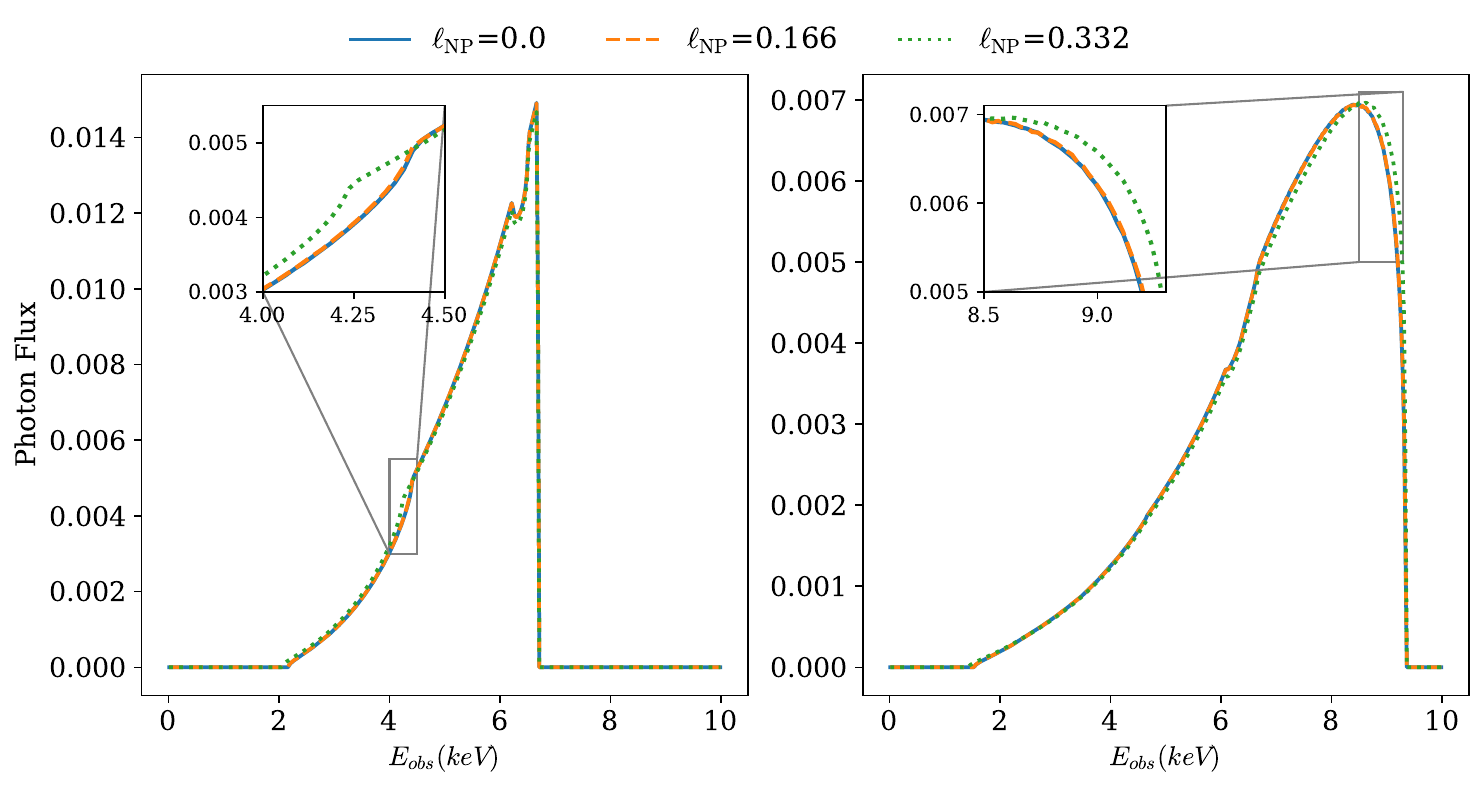}
\vspace{-0.2cm}
\caption{Impact of the deformation parameter $\ell_{\mathrm{NP}}$ on the iron line shape. The spacetime is described by the metric in Eqs.~\eqref{eqn:non_circular_metric} and~\eqref{eqn:non_circular_mass} with the spin parameter $a_*=0.8$. The left panel corresponds to a system with inclination angle $\iota=30^{\circ}$, while the right panel corresponds to $\iota=80^{\circ}$. Three cases with deformation parameters $\ell_{\mathrm{NP}}=0, 0.166, 0.332$ are plotted. Note that $\ell_{\mathrm{NP}}=0.332$ is the highest allowed value of the deformation parameter for $a_*=0.8$. 
\label{fig: ironlineplots}}
\end{figure*}

\begin{figure*}[t]
\centering
\includegraphics[width=0.9\linewidth]{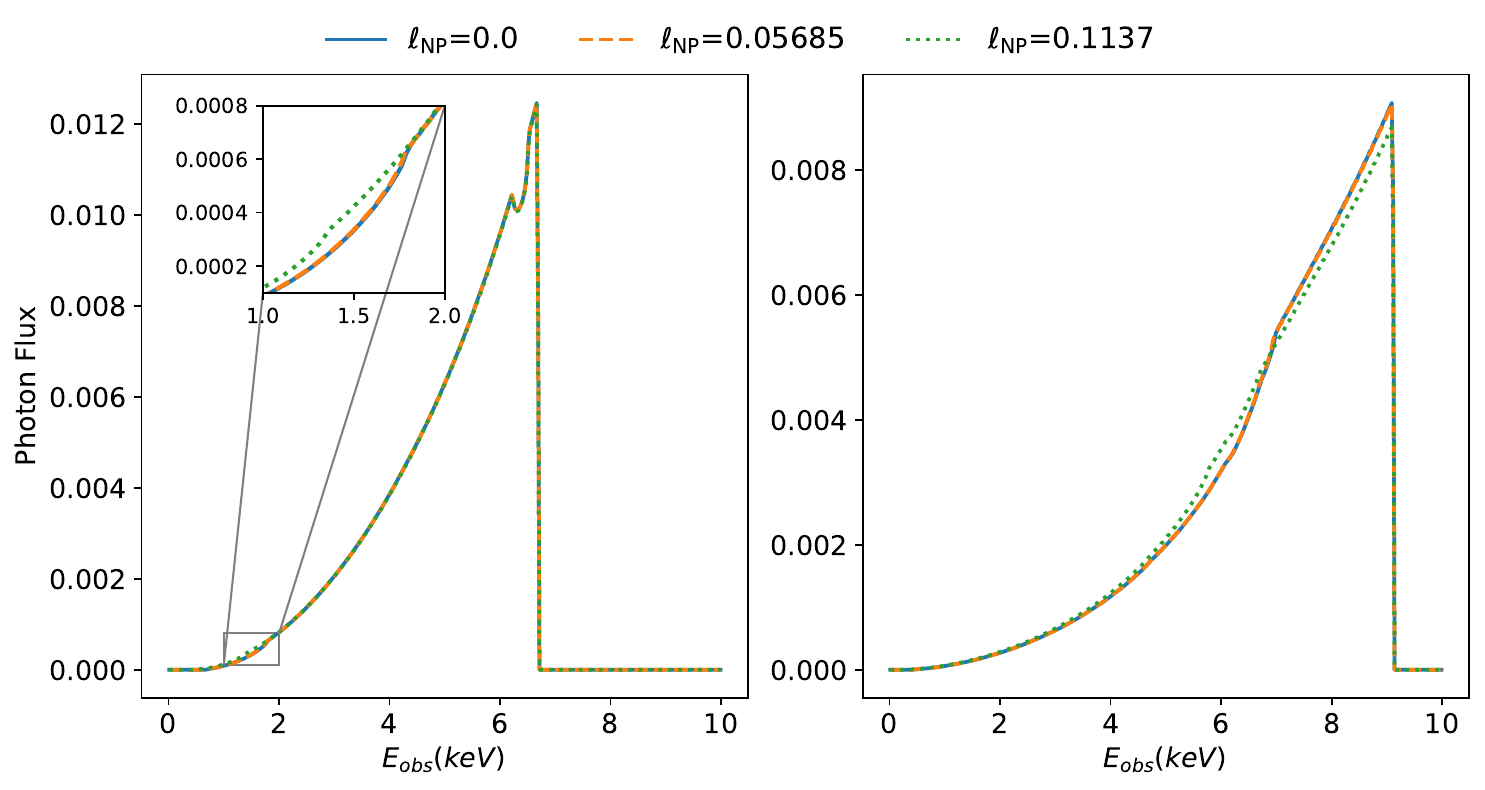}
\vspace{-0.2cm}
\caption{Same as Fig.~\ref{fig: ironlineplots}, but for the spin parameter $a_*=0.99$. Here, $\ell_{\mathrm{NP}}=0.1137$ is the highest allowed value of the deformation parameter.
\label{fig: ironlineplotsa99}}
\end{figure*}

In Fig.~\ref{fig: ironlineplots}, we consider cases with the spin parameter $a_*=0.8$, inclination angles $\iota=30^{\circ}$ and $80^{\circ}$, and deformation parameters $\ell_{\mathrm{NP}}=0, 0.166, 0.332$. The value $\ell_{\mathrm{NP}}=0.332$ represents the maximum allowed deformation parameter for $a_*=0.8$. 
Similar plots for the spin parameter $a_*=0.99$, inclination angles $\iota=30^{\circ}$ and $80^{\circ}$, and deformation parameters $\ell_{\mathrm{NP}}=0, 0.05685, 0.1137$ are shown in Fig.~\ref{fig: ironlineplotsa99}. 

We find that the high-inclination cases exhibit a stronger deformation effect, whereas the cases with inclination angle $\iota=30^{\circ}$ are almost identical to the Kerr spacetime. 
The deformation has only a minor impact on the shape of the broadened iron lines, even for the maximum deformation $\ell_{\mathrm{NP,max}}$. The profiles for $\ell_{\mathrm{NP}}=0$ and $\ell_{\mathrm{NP,max}}/2$ are nearly indistinguishable. This indicates that the deviations in the reflection spectrum do not grow linearly with the increase of the deformation parameter $\ell_{\mathrm{NP}}$; noticeable effects appear only when $\ell_{\mathrm{NP}}$ approaches its limit. These trends are also visible in the right panel of Fig.~\ref{fig:isco_non_circular}. The contour lines for the ISCO radius remain almost flat and then decrease rapidly as they approach the limit of deformation.

\section{OBSERVATIONAL CONSTRAINTS}~\label{sec:observationa_constraints}

In this section, we present the spectral analysis of a \textit{NuSTAR}~\cite{NuSTAR:2013yza} observation of the Galactic black hole EXO 1846--031 using the reflection model constructed in Sec.~\ref{sec: X-ray_reflection_spectroscopy} with the relativistic effect modeling with the non-circular metric introduced in Sec.~\ref{subsec: non-circular_metric}.

\subsection{Observations and data reduction}

EXO 1846--031 is a black hole X-ray binary first discovered by $\textit{EXOSAT}$ on April 3, 1985~\cite{1985IAUC.4051....1P}. After being in a quiescent state for about 34 years, a new outburst was first detected by $\textit{MAXI}/\textit{GSC}$ on July 23, 2019~\cite{2019ATel12968....1N}. The signal from EXO 1846--031 was detected by the X-ray mission $\textit{NuSTAR}$ on August 3, 2019 under the observation ID 90501334002~\cite{Draghis:2020ukh}. 
In this observation, this source showed a spectrum with strong reflection features, which is characterized by a prominent and broadened iron line around $7\,\mathrm{keV}$ and a Compton hump peaking around $20-30\,\mathrm{keV}$. Moreover, $\textit{NuSTAR}$ is the most suitable X-ray observatory for the analysis of reflection features of bright sources.
Therefore, this source has been widely used to test the reflection models~\cite{Abdikamalov:2021rty,Tripathi:2021wap,Li:2024eue} and spacetime metrics from modified gravity theories~\cite{Tripathi:2020yts,Tripathi:2021rwb,Yu:2021xen,Gu:2022grg,Tao:2023hou}. 

For the \textit{NuSTAR} observation of EXO 1846--031, we follow the procedures in Ref.~\cite{Li:2024eue} and reduced the data with NuSTARDAS and the CALDB 20220301~\cite{10.1117/1.JATIS.8.3.034003}. The source spectra is extracted from a circular region with a radius of $180^{\prime\prime}$ selected on both focal plane module A (FPMA) and focal plane module B (FPMB) detectors. A background region of comparable size was selected far from the source region for preventing influence from source photons. Afterwards, $\texttt{nuproducts}$ are used to generate the source and background spectra. The final spectra are grouped with the optimal binning algorithm in Ref.~\cite{Kaastra:2016qwt} by using the $\texttt{ftgrouppha}$.

\subsection{Data Analysis}

We use XSPEC v12.13.0c~\cite{1996ASPC..101...17A} to analyze the spectra. The model used here is $\texttt{constant*tbabs*(diskbb+nthcomp+relxillionCp\_nk)}$, as described in Sec.~\ref{subsec: X-ray_spectrum}. The best-fit values for this model and their uncertainties are listed in Table~\ref{tab:bestfit_new}. All values show good consistency with the similar model in Ref.~\cite{Li:2024eue}, which indicates the success of our modified relativistic model. The physical meanings of the fitting parameters are defined in Sec.~\ref{subsec: X-ray_spectrum}. For the \textit{NuSTAR} data, \texttt{constant} is a cross-calibration constant between FPMA and FPMB. During the analysis, the disk’s inner and outer edge parameters are set to the ISCO and $1000\,M$, respectively. Since the geometry of the corona is unknown, we model the emissivity profile using an empirical broken power-law ($\epsilon\propto1/r^{q_\textrm{in}}$ for $r<R_\textrm{break}$, $\epsilon\propto1/r^{q_\textrm{out}}$ for $r>R_\textrm{break}$) in the model \texttt{relxillionCp\_nk}. $A_\textrm{Fe}$ is the iron abundance in solar units, and $\log\xi$ represents the ionization parameter at the inner edge of the accretion disk.

\begin{table}
\caption{Best-fit values for Model $\texttt{constant*tbabs*(diskbb+nthcomp+relxillionCp\_nk)}$. The errors correspond to the 90\% confidence intervals. Parameters hitting limits are marked with $P$. }
\label{tab:bestfit_new}
\centering
\renewcommand{\arraystretch}{1.3}
\begin{tabular}{lc}
\toprule
\hline
\textbf{Model Parameter} & \textbf{Value} \\
\hline
\multicolumn{2}{l}{\texttt{constant}} \\
Factor & $1.015_{-0.001}^{+0.001}$ \\
\hline
\multicolumn{2}{l}{\texttt{tbabs}} \\
$N_{\rm H}$ [$10^{22}$ cm$^{-2}$] & $5.02_{-0.10}^{+0.21}$ \\
\hline
\multicolumn{2}{l}{\texttt{diskbb}} \\
$T_{\rm in}$ [keV] & $0.32_{-0.04}^{+0.05}$ \\
Norm & $1.58_{-0.19}^{+6.06} \times 10^{5}$ \\
\hline
\multicolumn{2}{l}{\texttt{nthComp}} \\
$\Gamma$ & $2.11_{-0.04}^{+0.04}$ \\
$kT_e$ [keV] & $58_{-5}^{+15}$ \\
Norm & $1.16_{-0.09}^{+0.02}$ \\
\hline
\multicolumn{2}{l}{\texttt{relxillionCp\_nk}} \\
$q_{\rm in}$ & $9.95_{-1.52}^{+P}$ \\
$q_{\rm out}$ & $0.00_{-P}^{+0.86}$ \\
$R_{\rm break}$ [$R_{\rm g}$] & $7.3_{-2.1}^{+3.1}$ \\
$a_*$ (Spin) & $0.982_{-0.003}^{+0.010}$ \\
$\iota$ [deg] & $76.4_{-2.3}^{+0.9}$ \\
$A_{\rm Fe}$ [Solar units] & $0.92_{-0.06}^{+0.45}$ \\
$\alpha_\xi$ & $0.18_{-0.03}^{+0.04}$ \\
$\log\xi\left[\rm erg~cm~s^{-1}\right]$ & $3.08_{-0.07}^{+0.09}$ \\
Norm & $5.91_{-0.32}^{+0.89} \times 10^{-3}$ \\
$\ell_{\mathrm{NP}}$ & $0.008_{-0.002}^{+0.025}$, $0.070_{-0.026}^{+0.014}$, $0.124_{-0.032}^{+P}$ \\

\hline
$\chi^2$/d.o.f. & $634.79/512$ \\
\hline
\bottomrule
\end{tabular}
\end{table}

\begin{figure}[t]
\centering
\includegraphics[width=0.9\linewidth]{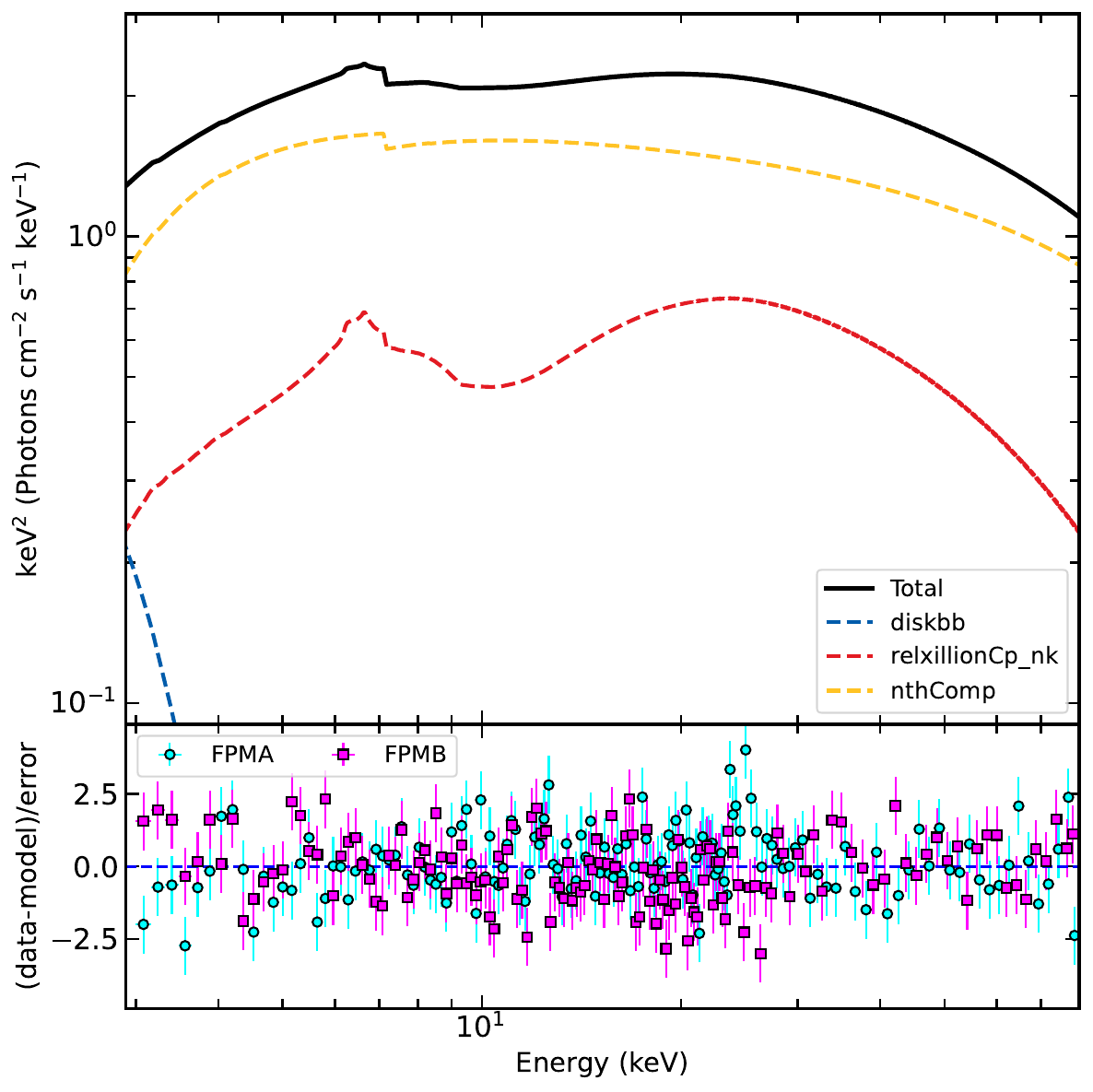}
\vspace{-0.2cm}
\caption{The upper panel shows the total best-fit model in black with the individual components: the thermal spectrum (\texttt{diskbb}) is blue-dashed line, the relativistic reflection spectrum (\texttt{relxillionCp\_nk}) is red-dashed line, and the Comptonized spectrum
from the corona (\texttt{nthComp}) is yellow-dashed line. The lower panel shows the data-to-best-fit model ratio, where the cyan and
magenta colors represent the data from the FPMA and FPMB sensors. Data is rebinned here for visual clarity. 
\label{fig: spectrum_components}}
\end{figure}

In the upper panel of Fig.~\ref{fig: spectrum_components}, we show the individual components of the spectrum for the best-fit model: the thermal spectrum \texttt{diskbb} is the dashed blue line, the relativistic reflection spectrum \texttt{relxillionCp\_nk} is the dashed red line, the Comptonized spectrum from the corona \texttt{nthComp} is the dashed yellow line, and the total spectrum is the solid black line. The lower panel shows the data-to-best-fit model ratio, where the cyan and magenta colors represent the data from the FPMA and FPMB sensors, respectively.

\begin{figure}[t]
\centering
\includegraphics[width=0.9\linewidth]{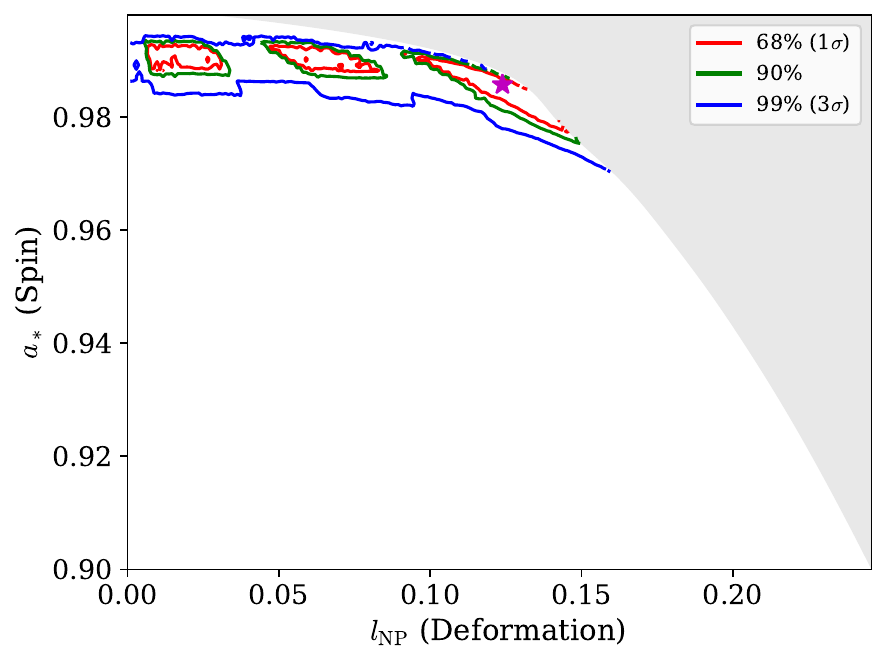}
\vspace{-0.2cm}
\caption{Constraints on the spin parameter $a_*$ and the deformation parameter $\ell_{\mathrm{NP}}$ obtained from the spectral fitting. The red, green, and blue contours correspond to the 68\% ($1\sigma$), 90\%, and 99\% ($3\sigma$) confidence intervals, respectively. The global minimum is marked by a magenta star. The gray region is excluded from our study because the spacetime possesses a naked singularity for these parameter combinations.
\label{fig: contour_non_circular}}
\end{figure}

In Fig.~\ref{fig: contour_non_circular}, we show the contour plot constraining the spin parameter $a_*$ and the deformation parameter $\ell_{\mathrm{NP}}$ derived from the fitting. In the plot, we show the contour lines for the 68\% ($1\sigma$), 90\%, and 99\% ($3\sigma$) confidence intervals in red, green, and blue, respectively. We find that there are three local minima in the fitting landscape, with a global minimum at $\ell_{\mathrm{NP}}=0.124$ indicated by a magenta star. The parameter values $\ell_{\mathrm{NP}}$ for these three local minima, along with their 90\% confidence intervals, are listed in Table~\ref{tab:bestfit_new}. In Sec.~\ref{subsec: single_line_shapes}, we showed the broadened iron line in the right panel of Fig.~\ref{fig: ironlineplotsa99} plotted with spin parameter $a_*=0.99$ and inclination angle $i=80^{\circ}$, which are close to their fitted values in Table~\ref{tab:bestfit_new}. 

However, we should not interpret a non-zero global minimum as tentative evidence for the existence of non-circular spacetime.
 First, the contour line of the 99\% confidence interval spans from the deformation parameter $\ell_{\mathrm{NP}} = 0$ to $\ell_{\mathrm{NP,max}}$.  Furthermore, as seen in the broadened iron line plots in Figs.~\ref{fig: ironlineplots} and \ref{fig: ironlineplotsa99}, the reflection spectrum is insensitive to this deformation even in the ideal case of high inclination and extremal deformation. Therefore, we expect that the search in the parameter space during fitting is insensitive to the parameter $\ell_{\mathrm{NP}}$. This flatness in $\chi^2$ space can easily result in an artificial minimum value from a numerical perspective. 
This suggests that the global minimum located at the near-extreme $\ell_{\mathrm{NP}}$, along with the other two local minima, may be artificial effects due to numerical fluctuations in the model grid.

\section{CONCLUSION}~\label{sec:conclusion}

In this work, we have presented the construction of a reflection model based on a specific form of non-circular metric. There has been growing interest in exploring the circularity of spacetime in recent years~\cite{Delaporte:2022acp,Ghosh:2024arw,Ghosh:2024het,Babichev:2025szb}. Although Ref.~\cite{Babichev:2025szb} proposed a general and elegant metric form to break the circularity condition, they did not provide a specific metric with strong physical motivation suitable for constructing a relativistic reflection model. We selected a metric constructed based on a locality principle proposed by Refs.~\cite{Eichhorn:2021iwq,Eichhorn:2021etc,Delaporte:2022acp}. This metric has several advantages. First, it breaks the circularity condition proposed in Refs.~\cite{Delaporte:2022acp,Babichev:2025szb} neatly by replacing the mass parameter in the Kerr metric with a mass function $M(r,\theta)$. Second, it is constructed with a physical implication based on a locality principle: the metric is deformed more significantly from the Kerr metric at locations with larger local curvature. Third, it possesses only one deformation parameter, which is crucial for constraining parameters through reflection spectrum analysis. Introducing multiple deformation parameters would make the construction of the FITS table for the transfer function computationally prohibitive and complicate the parameter space search during fitting. 

Horizon-penetrating coordinates, such as the ingoing Kerr coordinates used in Eq.~\eqref{eqn:non_circular_metric}, are favored when constructing non-circular metrics~\cite{Delaporte:2022acp,Babichev:2025szb}. Therefore, we derived the key equations and constructed the ray-tracing process for ingoing Kerr coordinates in Appendix~\ref{app: ray-tracing_ingoing}; this framework is not limited to our specific metric but is generally applicable. In Sec.~\ref{subsec: single_line_shapes}, the broadened iron line shapes are shown and discussed for different combinations of spin, inclination angle, and deformation parameter $\ell_{\mathrm{NP}}$. We find that more noticeable deformation in the high-energy part of the iron line appears at high inclination angles. Since high-energy photons originate from smaller radii as the inclination angle increases~\footnote{When the inclination angle is small and the Doppler effect is moderate, the emitted photons with the highest energies come from relatively large radii ($10\sim20\,M$). As the inclination angle increases, the Doppler effect becomes stronger while the gravitational redshift remains approximately the same. Therefore, the photons with the highest observed energy originate from smaller radii (approaching the inner side of the disk).}, this suggests that the effects of this deformation parameter are localized close to the black hole. This is consistent with the construction of the deformed metric: since the local curvature is larger near the black hole, the metric is more deformed in that region. Although a larger extremal deformation parameter is allowed for lower spin parameters, the extremal deformation in the low-spin case does not produce more significant deformation in the iron line shapes compared to the high-spin case. A possible reason is that the ISCO radius becomes smaller as spin increases, allowing photons to be emitted from smaller radii where the local curvature is larger and the metric is more strongly deformed. 

In this work, we analyzed the spectrum of the Galactic black hole X-ray binary EXO 1846--031 observed by \textit{NuSTAR} in 2019 using the model $\texttt{constant*tbabs*(diskbb+nthComp+relxillionCp\_nk)}$. EXO 1846--031 exhibits a strong reflection feature, a large inclination angle, and near-extremal spin, making it a suitable system for testing our relativistic model constructed from the deformed metric. As this is a well-studied source, the best-fit values for the model parameters show good consistency with Ref.~\cite{Li:2024eue}, including the spin parameter $a_*$ and inclination angle $\iota$. This serves as independent validation that our ray-tracing code in ingoing Kerr coordinates functions correctly.  
In Fig.~\ref{fig: contour_non_circular}, a global minimum is located at $a_*=0.982$ and $\ell_{\mathrm{NP}}=0.124$. However, we cannot constrain the deformation parameter $\ell_{\mathrm{NP}}$ strongly, as the 99\% ($3\sigma$) confidence interval spans the entire range from $\ell_{\mathrm{NP}} = 0$ to $\ell_{\mathrm{NP,max}}$ around the optimal spin region. Therefore, we do not claim that a non-zero $\ell_{\mathrm{NP}}$ is preferred by our model. Since this deformation effect already prefers a high-inclination system like EXO 1846--031, the constraint on $\ell_{\mathrm{NP}}$ in Fig.~\ref{fig: contour_non_circular} represents the likely limit of current capabilities using available electromagnetic data. To our knowledge, this is the first work to test a non-circular metric by analyzing the observational data of an electromagnetic spectrum emitted from an accreting black hole system. The generality of our analysis framework allows it to be easily applied to other types of non-circular metrics.


\vspace{0.5cm}

{\bf Acknowledgments --}
This work was supported by the National Natural Science Foundation of China (NSFC), Grant No.~W2531002.


\appendix
\section{Ray-tracing in ingoing Kerr coordinates}\label{app: ray-tracing_ingoing}
In this appendix, we explain how we implement the ray-tracing process in ingoing Kerr coordinates. First, we generate the photon initial conditions following the \textit{Relativistic Reflection Spectrum} section in Ref.~\cite{Bambi:2024hhi} using Boyer-Lindquist coordinates. To distinguish the $\phi$ coordinate in Boyer-Lindquist coordinates from that in ingoing Kerr coordinates, we denote the Boyer-Lindquist coordinates as $(t, r, \theta, \phi_{\mathrm{BL}})$. For a distant observer, the initial photon with position $(t_0, r_0, \theta_0, \phi_{\mathrm{BL},0})$ is generated with the initial 4-momentum $(k^t_0, k^r_0, k^\theta_0, k^{\phi_{\mathrm{BL}}}_0)$. The photon motion is characterized by two constants of motion: the specific energy $k_t=-E$ and the axial component of the specific angular momentum $k_{\phi_{\mathrm{BL}}}=L_z$.

We consider a metric of the form
\begin{equation}\label{eqn:ingoing_Kerr_coordinates}
\begin{split}
    \mathrm{d}s^2 &= g_{vv} \mathrm{d}v^2 + g_{\theta\theta} \mathrm{d}\theta^2 + g_{\phi\phi} \mathrm{d}\phi^2 \\
    &\quad + 2g_{v\phi} \mathrm{d}v \mathrm{d}\phi + 2g_{vr} \mathrm{d}v \mathrm{d}r + 2g_{r\phi} \mathrm{d}r \mathrm{d}\phi,
\end{split}
\end{equation}
where the metric components $g_{\mu\nu}$ for our case are given by the non-circular metric in Eq.~\eqref{eqn:non_circular_metric}.

The Lagrangian of a point-like free particle can be written as 
\begin{equation}
\begin{split}
    \mathcal{L} &= \frac{1}{2} \Big( g_{vv}\dot{v}^2 + 2g_{vr}\dot{v}\dot{r} + 2g_{v\phi}\dot{v}\dot{\phi} \\
    &\quad + 2g_{r\phi}\dot{r}\dot{\phi} + g_{\theta\theta}\dot{\theta}^2 + g_{\phi\phi}\dot{\phi}^2 \Big),
\end{split}
\end{equation}
where $\dot{} \equiv \mathrm{d}/\mathrm{d}\lambda$ and $\lambda$ is the affine parameter. We can find two constants of motion:
\begin{align}
    p_v &= \frac{\partial\mathcal{L}}{\partial\dot{v}} = g_{vv}\dot{v} + g_{vr}\dot{r} + g_{v\phi}\dot{\phi} = -E', \\
    p_\phi &= \frac{\partial\mathcal{L}}{\partial\dot{\phi}} = g_{v\phi}\dot{v} + g_{r\phi}\dot{r} + g_{\phi\phi}\dot{\phi} = L'_z,
\end{align}
where these two constants are named $E'$ and $L'_z$ in analogy to the constants $E$ and $L_z$ in Boyer-Lindquist coordinates.

For the deformed metric used here, the transformation between the coordinates $(v, \phi)$ in ingoing Kerr coordinates and $(t, \phi_{\mathrm{BL}})$ in Boyer-Lindquist coordinates for distant observer is given by
\begin{align}
    \mathrm{d}v &= \mathrm{d}t + \frac{r^2+a^2}{\Delta} \mathrm{d}r, \label{eqn:dv_transform_dt} \\
    \mathrm{d}\phi &= \mathrm{d}\phi_{\mathrm{BL}} + \frac{a}{\Delta} \mathrm{d}r, \label{eqn:dphi_transform_dphiBL}
\end{align}
where $\Delta := r^2 - 2Mr + a^2$ and $M$ is the mass parameter of the black hole.

By using Eqs.~\eqref{eqn:dv_transform_dt} and \eqref{eqn:dphi_transform_dphiBL}, we can transform the initial 4-momentum $(k^t_0, k^r_0, k^\theta_0, k^{\phi_{\mathrm{BL}}}_0)$ to $(k^v_0, k^r_0, k^\theta_0, k^\phi_0)$ in ingoing Kerr coordinates, where $k^v_0 = k^t_0 + (r^2+a^2)k^r_0/\Delta$ and $k^\phi_0 = k^{\phi_{\mathrm{BL}}}_0 + a k^r_0/\Delta$, while $k^r_0$ and $k^\theta_0$ remain unchanged. Therefore, the two new constants $E'$ and $L'_z$ can be obtained as
\begin{align}
    E' &= -k_v = -(g_{vv}k^v_0 + g_{vr}k^r_0 + g_{v\phi}k^\phi_0), \\
    L'_z &= k_\phi = g_{v\phi}k^v_0 + g_{\phi\phi}k^\phi_0 + g_{r\phi}k^r_0.
\end{align}

During the computation, a common choice is to use a normalized affine parameter $\lambda' = E\lambda$. When we solve for the photon trajectory along the geodesic, $\mathrm{d}v/\mathrm{d}\lambda'$ and $\mathrm{d}\phi/\mathrm{d}\lambda'$ can be calculated using two first-order equations derived from the conservation laws:
\begin{align}
    \frac{\mathrm{d}v}{\mathrm{d}\lambda'} &= \frac{-\frac{E'}{E}g_{\phi\phi} - \frac{L'_z}{E}g_{v\phi} - (g_{vr}g_{\phi\phi}-g_{v\phi}g_{r\phi})\frac{\mathrm{d}r}{\mathrm{d}\lambda'}}{g_{vv}g_{\phi\phi}-g_{v\phi}^2}, \\[1em]
    \frac{\mathrm{d}\phi}{\mathrm{d}\lambda'} &= \frac{\frac{E'}{E}g_{v\phi} + \frac{L'_z}{E}g_{vv} + (g_{vr}g_{v\phi}-g_{vv}g_{r\phi})\frac{\mathrm{d}r}{\mathrm{d}\lambda'}}{g_{vv}g_{\phi\phi}-g_{v\phi}^2}.
\end{align}

For the $r$ and $\theta$ coordinates, we solve the second-order differential equations, given in ingoing Kerr coordinates as:
\begin{widetext}
\begin{equation}
\begin{split}
    \frac{\mathrm{d}^2r}{\mathrm{d}\lambda'^2} =& -\Gamma^r_{vv}\left(\frac{\mathrm{d}v}{\mathrm{d}\lambda'}\right)^2 - \Gamma^r_{\phi\phi}\left(\frac{\mathrm{d}\phi}{\mathrm{d}\lambda'}\right)^2 - \Gamma^r_{\theta\theta}\left(\frac{\mathrm{d}\theta}{\mathrm{d}\lambda'}\right)^2 - 2\Gamma^r_{vr}\left(\frac{\mathrm{d}v}{\mathrm{d}\lambda'}\right)\left(\frac{\mathrm{d}r}{\mathrm{d}\lambda'}\right) \\
    &- 2\Gamma^r_{v\theta}\left(\frac{\mathrm{d}v}{\mathrm{d}\lambda'}\right)\left(\frac{\mathrm{d}\theta}{\mathrm{d}\lambda'}\right) - 2\Gamma^r_{v\phi}\left(\frac{\mathrm{d}v}{\mathrm{d}\lambda'}\right)\left(\frac{\mathrm{d}\phi}{\mathrm{d}\lambda'}\right) \\
    &- 2\Gamma^r_{r\theta}\left(\frac{\mathrm{d}r}{\mathrm{d}\lambda'}\right)\left(\frac{\mathrm{d}\theta}{\mathrm{d}\lambda'}\right) - 2\Gamma^r_{r\phi}\left(\frac{\mathrm{d}r}{\mathrm{d}\lambda'}\right)\left(\frac{\mathrm{d}\phi}{\mathrm{d}\lambda'}\right) - 2\Gamma^r_{\theta\phi}\left(\frac{\mathrm{d}\theta}{\mathrm{d}\lambda'}\right)\left(\frac{\mathrm{d}\phi}{\mathrm{d}\lambda'}\right),
\end{split}
\end{equation}
\end{widetext}
\begin{widetext}
\begin{equation}
\begin{split}
    \frac{\mathrm{d}^2\theta}{\mathrm{d}\lambda'^2} =& -\Gamma^\theta_{vv}\left(\frac{\mathrm{d}v}{\mathrm{d}\lambda'}\right)^2 - \Gamma^\theta_{\phi\phi}\left(\frac{\mathrm{d}\phi}{\mathrm{d}\lambda'}\right)^2 - \Gamma^\theta_{\theta\theta}\left(\frac{\mathrm{d}\theta}{\mathrm{d}\lambda'}\right)^2 \\
    &- 2\Gamma^\theta_{v\phi}\left(\frac{\mathrm{d}v}{\mathrm{d}\lambda'}\right)\left(\frac{\mathrm{d}\phi}{\mathrm{d}\lambda'}\right) - 2\Gamma^\theta_{r\theta}\left(\frac{\mathrm{d}r}{\mathrm{d}\lambda'}\right)\left(\frac{\mathrm{d}\theta}{\mathrm{d}\lambda'}\right) - 2\Gamma^\theta_{r\phi}\left(\frac{\mathrm{d}r}{\mathrm{d}\lambda'}\right)\left(\frac{\mathrm{d}\phi}{\mathrm{d}\lambda'}\right),
\end{split}
\end{equation}
\end{widetext}
where $\Gamma^\mu_{\nu\rho}$ are the Christoffel symbols of the metric in Eq.~\eqref{eqn:non_circular_metric}.

After the photon hits the accretion disk, the redshift factor and emission angle need to be calculated. We found that the final forms of the redshift factor and emission angle in ingoing Kerr coordinates are similar to their expressions in Ref.~\cite{Bambi:2024hhi} in Boyer-Lindquist coordinates. Therefore, we omit most derivation details and only show the key equations here.
The motion of the material in the disk can be approximated by quasi-geodesic, equatorial, circular orbits. The angular velocity of the material measured in the coordinates $(v,r,\theta,\phi)$ is 
\begin{equation}
    \Omega_K = \frac{\mathrm{d}\phi}{\mathrm{d}v} = \frac{- \partial_r g_{v\phi} + \sqrt{(\partial_r g_{v\phi})^2 - (\partial_r g_{vv})(\partial_r g_{\phi\phi})}}{\partial_r g_{\phi\phi}},
\end{equation}
where we consider co-rotating orbits. For the motion of the material, we have 
\begin{equation}
    \dot{v} = \frac{\mathrm{d}v}{\mathrm{d}\tau} = \frac{1}{\sqrt{-g_{vv} - 2g_{v\phi}\Omega_K - g_{\phi\phi}\Omega_K^2}}.
\end{equation}
The redshift factor $g$ can be computed with the expression
\begin{equation}
    g = \frac{E_{\mathrm{o}}}{E_{\mathrm{e}}}  = \frac{-E}{\dot{v}k_v + \Omega_K\dot{v}k_\phi},
\end{equation}
where $E_{\mathrm{o}}$ and $E_{\mathrm{e}}$ are the energy of the photon measured at the absorpotion point and the emission point and E is simply the conserved specific energy $E = -k_t$ of the photon. 

The emission angle $\vartheta_{\mathrm{e}}$ is given by
\begin{equation}
    \cos\vartheta_{\mathrm{e}} = \frac{\sqrt{g_{\theta\theta}}gk^\theta_f}{k_t} = -\frac{\sqrt{g_{\theta\theta}}gk^\theta_f}{E},
\end{equation}
where $k^\theta_f$ is the value of $k^\theta$ obtained when the photon hits the disk after backward ray-tracing from the distant observer.

\bibliography{NonCircular.bib}


\end{document}